\documentclass[onecolumn,showpacs,showkeys,preprintnumbers,amsmath,amssymb]{revtex4}

\usepackage{amsmath, amssymb, latexsym, verbatim}
\usepackage{graphicx}
\usepackage{dcolumn}
\usepackage{bm}
\usepackage[dvips]{color}
\usepackage[usenames,dvipsnames]{xcolor}

\newcommand{\nin}{\noindent}

\begin{document}

\title{A Quasi-Newtonian Approach to Bohmian Mechanics I: Quantum Potential}

\author{Mahdi Atiq}
 \altaffiliation{Sharif University of Technology, Physics Department, Tehran, Iran}
 \email{mma\_atiq@yahoo.com}

\author{Mozafar Karamian}
 \email{karamian@ymail.com}

\author{Mehdi Golshani}
 \altaffiliation{Institutes for Theoretical Physics and Mathematics (IPM), Tehran, Iran}
 \altaffiliation{Department of Physics, Sharif University of Technology, Tehran,
        Iran}
 \email{mehdigolshani@yahoo.com}

\begin{abstract}
\nin In this article, we investigate Bohm's view of quantum
theory, especially Bohm's quantum potential, from a new
perspective. We develop a quasi-Newtonian approach to Bohmian
mechanics. We show that to arrive at Bohmian formulation of
quantum mechanics, there is no necessity to start from the
Schr\"odinger equation. We also obtain an equation that restricts
the possible forms of quantum potential and determines the
functional form of it without appealing to the wave function and
the Schr\"odinger equation. Finally, we discuss about the
significance of quantum potential in the conceptual structure of
quantum theory.
\end{abstract}



\maketitle

\emph{Published in Annales de la Fondation Louis de Broglie, 2009}

\section{Introduction} \label{S:Intro}
\nin Quantum theory was developed in 1920's and it explained a
multitude of phenomena, including the atomic spectra. It
introduced concepts like \emph{wave function}, \emph{operators},
and \emph{eigenvalues}, but at the same time it undermined some of
the well-cherished philosophical principles like \emph{causality}.
In his 1952 papers \cite{Bo52}, David Bohm introduced a
formulation of quantum theory that kept many concepts of the
standard quantum mechanics and yielded the same empirical results,
but was \emph{causal}. In particular, Bohm introduced the concept
of \emph{quantum potential}, which could be taken as the source of
quantum novelties.\\

In this article we deal with the meaning and role of quantum
potential in the quantum theory. We start from a quasi-Newtonian
approach and then we show that by introducing the 'quantum
potential' concept into the mechanics of particles, one can get
the mathematical form of quantum potential by imposing the
requirement that the total energy of ensemble be minimized. Also
we show that one can get Bohm's basic equations without appealing
to the \emph{Schr\"odinger equation} and \emph{wave function}.

\section{Bohmian interpretation of quantum mechanics} \label{S:BM Approach}
\nin According to quantum mechanics, the time development of the
wave function of a one-particle system is described by the
time-dependent Schr\"odinger equation:

\begin{equation} \label{E:Shrod Eq}
i\hbar \frac{\partial}{\partial t} \psi=-\frac{\hbar^2}{2m}
\nabla^2 \psi+V \psi.
\end{equation}

\noindent Using Bohm's suggestion \cite{Bo52}, we write the
complex wave function $\psi$ in the polar form:
\begin{equation} \label{E:Psi Def}
\psi (x, t) = R (x, t) \, \exp (i S(x, t)/\hbar)
\end{equation}

\noindent in which $R\geq0$. Replacing equation \eqref{E:Psi Def}
into \eqref{E:Shrod Eq}, the complex equation \eqref{E:Shrod Eq}
reduces to the following real equations:

\begin{equation} \label{E:QHJ Eq}
\frac{ {( \nabla S )}^2 }{2m} + V(x)- \frac{\hbar^2}{2m}
\frac{\nabla^2 R}{R}+ \frac{\partial S}{\partial t}=0
\end{equation}

\begin{equation} \label{E:Contin Eq}
\frac{\partial R^2}{\partial t}+ \nabla . ( R^2 \frac{\nabla S}{m}
)=0.
\end{equation}

\noindent The equation \eqref{E:Contin Eq} is the so-called
continuity equation, which is written in terms of $R$ and $S$. The
equation \eqref{E:QHJ Eq} is very similar to the Hamilton-Jacobi
equation of classical mechanics. Therefore, Bohm \cite{Bo52}
suggested that we take $S$ as Hamilton's principal function and
take the momentum and energy of the particle to be

\begin{equation} \label{E:E & P Def}
p=\nabla S, \quad E=-\frac{\partial S}{\partial t}.
\end{equation}

In this view, we can discuss about the path of the particle, like
in classical mechanics, something not permissible in the
Copenhagen interpretation of quantum mechanics. Since $S$
describes the phase of the wave function, and in quantum mechanics
the wave function is taken to be single-valued, thus $S$  has to
be single-valued, apart from an additive constant such as $2\pi n
\hbar$ ($n$ is integer). $R$  is also taken to be single-valued.
While $S$ in the classical mechanics is a multi-valued auxiliary
function \cite[Sec. 2.2.2]{Ho93}, \cite[Chap. 10]{Go02}, in the
\emph{ ordinary} Bohmian interpretation, it seems that, it has
more share of physical reality, relative to the classical case. In
other words, in quantum mechanics, apparently, $S$ has a role in
the dynamics of the particle \cite[Chap. 3]{Ho93}. The expression

\begin{equation} \label{E:BQP Def}
Q(x)=-\frac{\hbar^2}{2m}\frac{\nabla^2 R}{R}
\end{equation}

\noindent in equation \eqref{E:QHJ Eq} is called quantum
potential, and in the Bohmian interpretation it can explain the
non-classical behaviors of particles, such as interference,
barrier penetration, etc. In short, we can say that in the usual
Bohmian interpretation, the particle is under the influence of $R$
and $S$, in addition to the external potential $V(x)$. In this
interpretation, one assumes the fundamental Schr\"odinger
equation, but tries to extract another meaning from the wave
function. The Bohmian mechanics, as we know it, is not usually
taken to be a theory independent from the standard quantum theory.
But it attempts to dispense with some of the interpretational
aspects of quantum theory, such as indeterminism and the lack of
particle trajectory. Like ordinary quantum mechanics, the
fundamental element of the ordinary Bohmian mechanics is the wave
function which develops according to the Schr\"odinger equation.
But the phase and the amplitude of the wave function are
interpreted in such a way that the concept of particle and its
path remain intact. Even, if one finds cases which the predictions
of the Bohmian mechanics and the ordinary quantum mechanics are
different, this does not mean that we are dealing with two
basically different theories. Because, when we interpret the
elements of a theory (e.x. position and momentum) in a special
manner and give special meaning to those elements, the method of
problem solving, and consequently, the predictions, could be
affected. Therefore, that interpretation will be adopted which
fits the empirical results better.

\section{Quasi-Newtonian approach}\label{S:New Approach}
\nin If we look at the ideas of David Bohm
\cite{Bo52}, \cite{Bo80} we find that, his main purpose to develop
the so-called Bohmian interpretation was to prove that the von
Newmann's argument about the impossibility of describing the
current quantum mechanics on the basis of 'Hidden variables' is
wrong. He realized that by supposing a strictly well-defined
localized particle with a well-defined trajectory that coexists
with the wave and interpreting $\nabla S$ and $-\partial S /
\partial t$ as momentum and energy of the particle that the
Schr\"odinger wave function describes, we can consistently
describe all known quantum phenomena. But, what he found was very
powerful, specially in solving the measurement problem in the
quantum mechanics \cite{Bo52}, \cite[chap. 8]{Ho93}. In the
Bohmian interpretation it is very simple to show how the
measurement process terminates with one of the eigenvalues of
relevant quantum mechanical operator \cite[chap. 8]{Ho93}, without
needing the collapse of the wave function. Therefore, one of the
merits of including a localized particle with well-defined
trajectory to the quantum theory is to reducing the number of
postulates we need to describe quantum phenomena. This fact by
itself is sufficient to show that the Bohm's trajectories are not
some artificial curves added to a pre-existing quantum theory. In
a subsequent paper we show that, even there is no need for
postulating the 'eigenvalue postulate'. It is a natural
consequence of Bohmian approach to \emph{prove} that even prior to
any measurement the energy and angular momentum of electron in the
atom in stationary states are eigenvalues of relevant operators.

What is revolutionary, in the Bohm interpretation, in contrast to
the Copenhagen interpretation of quantum mechanics? The answer
absolutely is, 'casuality' and 'trajectories'. In the Copenhagen
interpretation, the only casual element is the evolution of wave
function with time. There is no other casual element in that
interpretation. Because of the lack of casuality, the particles do
not have any well-defined trajectories in space and time. In this
respect, the Bohm's idea is revolutionary. In other words, in
comparison with formulation of orthodox quantum mechanics, the
Bohm's idea is revolutionary. In this viewpoint, the main part of
Bohmian formulation of quantum mechanics is the equation $p=\nabla
S$, by which we can \emph{define} or \emph{obtain} the particle's
trajectory. According to some authors, for interpreting $\nabla S$
as particle momentum, appealing to the modified Hamilton-Jacobi
equation \eqref{E:QHJ Eq} is not a necessity. One can arrive at

\begin{equation}\label{E:V = del S = del psi on psi}
    \dot{x}=\frac{\hbar}{m} Im \frac{\nabla \psi}{\psi}=\frac{\nabla S}{m}
\end{equation}

\noindent only by appealing to some symmetry arguments about the
wave function itself \cite{DGZ92},\cite{DGZ96}. In this view, the
only property of Bohm's particle is its position. This position
evolves with time, according to \eqref{E:V = del S = del psi on
psi}. The wave function is responsible for the evolution of
particle's position with time. It seems that, according to this
viewpoint, there is no need for the Newtonian concepts such as
energy, momentum, angular momentum, $\cdots$, for the particle.
The particle has only a position that the evolution of which is
determines strictly by the wave function.

As we mentioned, if we compare the Bohmian mechanics with the
Copenhagen interpretation, the new revolutionary elements are the
'casuality' and 'trajectories'. But, just these two elements are
'trivial' and 'no revolutionary', if we compare the Bohmian
mechanics with classical (Newtonian or Einsteinian) mechanics. We
should not forget that the classical mechanics is successful
theory of the world around us. The presence of 'quantum phenomena'
must not be the cause of disregarding the 'classical phenomena',
such as the freedom of throwing a thing with arbitrary initial
momentum, and the absence of interference phenomenon for
particles.

Therefore, in comparison with classical mechanics, the main
element of Bohmian mechanics is not the equation $p=\nabla S$.
Indeed, whenever we have a Newtonian force equation like $dp / dt
= - \nabla V$ or $dp / dt = -\nabla(V+Q)$(as we have from Bohm's
formulation), we can find a $S$ function such that we have
$p=\nabla S$. This equation is a mathematical definition. Indeed,
arriving at this relation by starting from a Newtonian force law
is much simpler that arriving at it through symmetry
considerations about the wave function. In this regard, the
revolutionary parts of Bohmian mechanics are the presence of a
highly non-classical potential, named quantum potential, and the
Born rule -issues that are not independent, because both depend on
function $R$. In the classical mechanics, there is neither quantum
potential, nor the Born rule.

In this paper and subsequent papers, we try to develop a
quasi-Newtonian approach to Bohmian mechanics.

\subsection{Illustrating the quasi-Newtonian approach} \label{S:Illus. quasi-Newtonian picture}
\nin In the quasi-Newtonian approach, we try to describe quantum
phenomena in a manner nearest to the classical mechanics. Although
the quantum phenomena are non-classical, we encounter them as some
new regularities which had been hidden. They are rules of nature
which do not make manifest themselves in the daily experiences.

One of the most important quantum phenomena is two slit
interference. When a flux of identical particles passes from two
slits and reaches a screen, it shows a wave like dark-bright
fringes, in the sense that the number density of particles on the
screen differs from bright regions to dark regions. If we emit the
particles one by one on the slits, after a long time the
dark-bright pattern appear. This fact shows that the whole pattern
is the result of the behavior of individual particles. If we
believe that the behavior of individual particles in this
experiments is essentially deterministic, we expect, in a
quasi-Newtonian picture,  that an extra potential must be
responsible for this novel behaviors. The effect of this extra
potential is such that prevents the particles to fall in dark
regions and forces them to fall almost in the central regions of
bright fringes.

If we denote the number density of the particles by $\rho$, and
the apparent extra potential that acts on the individual particles
by $Q$, and try to denote the mathematically unknown fundamental
agent of these novel behaviors by $\chi$ we must have

\begin{equation}\label{E:rho depend chi}
    \rho = f(\chi)
\end{equation}

\begin{equation}\label{E:Q depend chi}
    Q = Q(\chi).
\end{equation}

But the Eq. \eqref{E:rho depend chi} can not specifies $\rho$
uniquely, because we can emit a flux with arbitrary intensity.
Therefore, the total number of particles in the flux is arbitrary.
Indeed, this equation must specifies the form of distribution of
$\rho$ in the space and on the screen. We can fix the exact values
of $\rho$ by specifying the value of $\rho$ at a given point $x_0$
as $\rho_0$. Inserting this condition into \eqref{E:rho depend
chi}, it must specify $\rho$ uniquely. Another way for specifying
$\rho$ uniquely, is to specify the total number $N$ of particles
in the flux. This is accomplished by the relation

\begin{equation}\label{E:rho normalization by N}
    \int \rho \, d^3 x = N.
\end{equation}

\noindent Dividing both sides of this relation by $N$, we can
always obtain a normalized $\rho$ such that

\begin{equation}\label{E:rho normalization}
    \int \rho \, d^3 x = 1.
\end{equation}

Suppose that we can mathematically get the inverse of the Eq.
\eqref{E:rho depend chi} in the form:

\begin{equation}\label{E:chi depend rho}
    \chi = f^{-1}(\rho).
\end{equation}

\noindent By this relation we can eliminate the unknown agent
$\chi$ from the problem. Inserting this equation into \eqref{E:Q
depend chi} we obtain

\begin{equation}\label{E:Q depend rho}
    Q = Q(\rho).
\end{equation}

Note that the dependence of $Q$ on $\rho$ is not basically direct
and is only a technic for solving the problem, by eliminating the
unknown factor $\chi$.

Conforming to the usual Bohmian mechanics formulation, we
introduce a function $R$ such that we have

\begin{equation}\label{E:rho = R2}
    \rho = R^2.
\end{equation}

According to the usual formulation of Bohmian mechanics, one may
think of $R$ as being the amplitude of a wave function. But, we
don't have any emphasis on $R$ as being the amplitude of a wave
function, and we can even show that this is not a good
interpretation, because it would be suitable to allow negative
values for $R$.

Due to \eqref{E:rho = R2} we can rewrite Eq. \eqref{E:Q depend
rho} as

\begin{equation}\label{E:Q depend R}
    Q=Q(R).
\end{equation}

\noindent This equation means that $Q$ could be function of $R$
and its partial derivatives.

Now, We consider the Hamiltonian of a single particle in three
dimensions:

\begin{equation} \label{E:Hamiltonian}
H(x,p, t)=\frac{p^2}{2m}+V(x)+Q(R(x, t))
\end{equation}

\noindent where the quantum potential $Q$ is taken to be unknown
function of $R$. For the moment, we assume that the function $R$
does not depend on time and therefore the energy is conserved in
the presence of quantum potential $Q$ . Therefore, the energy of
particle is taken to be

\begin{equation} \label{E:Particle Energy}
E=H(x, p)=\frac{p^2}{2m}+V(x)+Q(R(x)).
\end{equation}

Without knowing the functional dependence of $Q$ with respect to
$R$ we are not able to solve the problem of finding the path of
the particle. We need extra assumptions about $R$ and $Q$. We
appeal to a simple (or the simplest) assumption: the total energy
of the ensemble of particles must be minimized.

This means that we minimize the integration

\begin{equation}\label{E:Total E int}
    \int \rho \, H d^3 x
\end{equation}

\noindent while keeping the condition \eqref{E:rho normalization}.
According to variations calculus, we can write this requirements
as

\begin{equation}\label{E:Var principle}
   \delta \int \rho \left \{ H - \lambda \right \} d^3 x = 0
\end{equation}

\noindent in which $\lambda$ is Lagrange's undetermined
multiplier.

This equation is not useful unless we could write Hamiltonian
completely as a function of space coordinates only. This is
feasible by using Hamilton-Jacobi's principal function $S$.
According to the Hamilton-Jacobi theory, we can express the
momentum of particle as

\begin{equation}\label{E:p = del S}
    p = \nabla S
\end{equation}

\noindent and for conserved systems we have

\begin{equation}\label{E:S=W-Et}
    S(x, t) = W(x)-Et.
\end{equation}

Therefore, using the Eq. \eqref{E:p = del S}, we can express the
Hamiltonian as a function of space coordinates only and rewrite
Eq. \eqref{E:Var principle} as

\begin{equation}\label{E:Var principle2}
   \delta \int R^2 \left \{ \frac{(\nabla S)^2}{2m} + V + Q - \lambda \right \} d^3 x =
   0.
\end{equation}

This equation is an eigenvalue problem with $\lambda$ as
eigenvalue. Therefore we call $\lambda$ the energy eigenvalue. It
is simple to prove that whenever the quantum potential $Q$ is in
the Bohmian form \eqref{E:BQP Def}, $\lambda$ is identical to
eigenvalue of the time-independent Schr\"odinger equation
(Appendix). In other words, the Eq. \eqref{E:Var principle2} is an
integral form of the energy eigenvalues differential equation.

For the reasons, which will be explained later, we assume $Q$ to
be a function of $R$ and its first and second derivatives. Indeed,
we show that the insufficiency of first order derivatives and  the
presence of second derivatives is a necessity for the existence of
non-trivial quantum potential.

If we denote the integrand of \eqref{E:Var principle2} by $g$,
using summation rule for indices $i$, $j$ and abbreviation
$\partial_i$ for the partial derivative $\partial /
\partial x_i$, and so on, we have from variational calculus

\begin{equation} \label{E:g-R}
\frac{\partial g}{\partial R}-{\partial_i}( \frac{\partial
g}{\partial ( {\partial_i R} )} )+{\partial_i}{\partial_j}(
\frac{\partial g}{\partial ( {\partial_i}{\partial_j R} )} )=0
\end{equation}

\begin{equation} \label{E:g-S}
{\partial_i}( \frac{\partial g}{\partial ( {\partial_i S} )} )=0.
\end{equation}

From the equation \eqref{E:g-R} it follows that:

\begin{equation} \label{E:g-R Expanded}
2R \Big\{  \frac{ {( \nabla S )}^2 }{2m} +V+Q-\lambda \Big\} + R^2
\frac{\partial Q}{\partial R}-{\partial_i}( R^2 \frac{\partial
Q}{\partial ( {\partial_i R} )} )+{\partial_i}{\partial_j} ( R^2
\frac{\partial Q}{\partial ( {\partial_i}{\partial_j R} )} )=0.
\end{equation}

The expression between braces is $E-\lambda$. Therefore, this
equation reduces to

\begin{equation} \label{E:Q Condition Lambda-E}
R^2 \frac{\partial Q}{\partial R}-{\partial_i} ( R^2
\frac{\partial Q}{\partial ( {\partial_i R} )} )+
{\partial_i}{\partial_j} ( R^2 \frac{\partial Q}{\partial (
{\partial_i}{\partial_j R} )} )=\\
2R \{ \lambda - E \}.
\end{equation}

The quantity $E$ is the particle energy, i.e., is a constant
related to the particle dynamics. On the other hand, the energy
eigenvalue $\lambda$ is a constant related to the particle
dynamics. Thus, it is natural to take them to be identical.
Indeed, we seek cases where the energy of the particle is equal to
the energy eigenvalue. Theoretically, there is the possibility for
the particle energy to be different from energy eigenvalues. But,
we take them identical here. Therefore, we have

\begin{equation} \label{E:Lambda=E=??}
\lambda= E= \frac{ {( \nabla S ) }^2 }{2m}+V(x)+Q
\end{equation}

\begin{equation} \label{E:Q Condition}
R^2 \frac{\partial Q}{\partial R}-{\partial_i} ( R^2
\frac{\partial Q}{\partial ( {\partial_i R} )} )+
{\partial_i}{\partial_j} ( R^2 \frac{\partial Q}{\partial (
{\partial_i}{\partial_j R} )} )=0.
\end{equation}

Consequently, the equation \eqref{E:Q Condition} is an important
condition that quantum potential $Q$ must fulfill. From the
equation \eqref{E:g-S}, one gets

\begin{equation} \label{E:Contin Eq -Stationary}
\nabla . ( R^2 \frac{\nabla S}{m} )=0
\end{equation}

\noindent which is the so-called continuity equation for
stationary states.

\subsection{The derivation of the quantum potential} \label{S:Derivating Quantum Potential}
\nin We have to find a function $Q(R)$ which satisfies the
equation \eqref{E:Q Condition}. The form of $Q$ with respect to
$R$ must be such that the Eq. \eqref{E:Q Condition} is satisfied
for every arbitrary $R$. Indeed, we do not have inclination that
the condition \eqref{E:Q Condition} restricts the acceptable forms
of $R$. Our interest is to restrict the functional form of $Q$
with respect to $R$, not the form of $R$ with respect to $x$. The
simplest solution is

\begin{equation} \label{E:Q=const}
Q \equiv const
\end{equation}

\noindent in which, $Q$ appears as an additive constant in the
energy equation. This is satisfied for arbitrary $R$. If we did
not have the equation \eqref{E:Contin Eq -Stationary}, we could
obtain from equation \eqref{E:Q=const} the whole of classical
mechanics, apart from a constant value in the energy. The
condition \eqref{E:Contin Eq -Stationary} imposes some
restrictions on the particle motion (such as prevention of turning
points in the particle orbits, as we shall see in a subsequent
paper) -conditions which are not imposed in the classical
mechanics. Therefore, if we restrict ourselves to stationary
states we can not describe the whole classical dynamics. To
describe the whole classical dynamics we should consider equation
\eqref{E:Q=const} for non-stationary states. If the constant $Q$
is non-zero, we should not, however, confuse this constant with
the origin of potential energy in classical mechanics, as we have
not made any changes in the origin of potential. Thus, a constant
$Q$ would be a real non-classical term. If it is non-zero and
dependent on the particle, one may interpret it as the rest
energy, but there is no way to prove
this.\\
\indent Now, we can expect that more complicated forms of $Q$
would lead to non-classical results. We are looking for a
non-trivial expression for $Q$. Consider that, $Q$ is a function
of $R$ and its first and second derivatives. We shall see a little
later that the first-order derivatives are not sufficient for
getting a non-trivial quantum potential. The quantum potential $Q$
is a scalar function and therefore must be rotational-invariant.
Thus, we expect that the first and second derivatives of $R$
appear in the form of $\left| \nabla R \right|$ and $\nabla^2 R$,
respectively. Therefore, $Q$ is constructed from the factors $f_1
= R^m$, $f_2 = \left| \nabla R \right|^n$ and $f_3 = ( \nabla^2 R
)^p$ for some unknown powers $m$, $n$ and $p$. Among all the
expressions that one can write by summation or multiplication of
these factors the only expression that leads to a non-trivial form
for quantum potential is $f_1 f_2 f_3$, i.e.

\begin{equation} \label{E:General form of Q}
Q=A \, R^m \left| \nabla R \right|^n ( \nabla^2 R )^p.
\end{equation}

We emphasize that \emph{only expressions in the form of
\eqref{E:General form of Q} lead to a non-trivial solution for
equation \eqref{E:Q Condition}}. Inserting the equation
\eqref{E:General form of Q} into \eqref{E:Q Condition}, one can
show, after some elementary (but, to some extent long)
calculations, that only two sets of values for $m$ , $n$ and $p$
can lead to a satisfactory solution for the equation \eqref{E:Q
Condition}:

\begin{align}
m&=0, \quad n=0, \quad p=0 \nonumber \\
m&=-1, \quad n=0, \quad p=1. \nonumber
\end{align}

In the first case, we get trivial solution $Q(x)=A=const$ , which
we have already discussed. The second case leads to the result
that we have in ordinary Bohmian mechanics. We observe that with
$p=0$ , the equation \eqref{E:Q Condition} can be satisfied for no
values other than zero for $m$ and $n$. Remembering that no
expressions other than \eqref{E:General form of Q} can lead to a
non-trivial solution for $Q$, we observe that the presence of a
non-trivial solution for $Q$ requires $p \neq 0$, i.e., it shows
the necessity of second-order derivatives.

\indent Thus, the simplest non-trivial form of the quantum
potential is in the form

\begin{equation} \label{E:BQP with A}
Q(x)=A \frac{\nabla^2 R}{R}.
\end{equation}

This means that we not only got the quantum Hamilton-Jacobi
equation and the continuity equation, but we also justified the
form of $Q$ in terms of $R$. Thus, if we are to have a quantum
potential, its simplest non-trivial form is the familiar one. We
observe that the form of quantum potential \eqref{E:BQP with A} is
a mathematical necessity for minimizing the total energy of the
ensemble rather than being a consequence of the Schr\"odinger
equation. This shows the power of quantum potential concept in the
quantum theory.\\

The constant value of $A$ and specifically its sign in the
equation \eqref{E:BQP with A} are significant. Any departure from
the value of $A$ that we get from quantum mechanics leads to
serious changes in the particle dynamics. But, here we don't have
any independent way for getting its value. It seems that the
simplest way for obtaining the constant $A$ is by adapting the
energy levels of Hydrogen atom in the theory with those obtained
from Bohr's model. This is exactly what Schr\"odinger did in his
original works for finding some constants \cite[p. 8]{Sch28}. We
expect that this method yields the value $-\hbar^2/2m$ for $A$,
and therefore we take it simply to be $-\hbar^2/2m$.\\
\indent Note that in this discussion we have not made any use of
the concept of wave function. Here $S$ is a mathematical function,
the derivatives of which gives momentum and energy, and $R$ is
representative of a new physical entity which contributes to the
dynamics of the particle through $Q$.

This can means that quantum potential is a more fundamental
concept than wave function and Schr\"odinger equation.\\

\indent If the state is not stationary, i.e., $R$ depends on time
and $S$ is a general function of time and space, we can directly
use

\begin{equation} \label{E:Int R2 H+St}
\delta \int R^2 \Big\{ H(x, S(x, t), R(x, t)) + \frac{\partial
S}{\partial t} \Big\} \, d^4 x=0
\end{equation}

\noindent and the relations \eqref{E:g-R}, \eqref{E:g-S} and
\eqref{E:BQP with A}, to obtain the Hamilton-Jacobi and continuity
equations (here the index $i$ includes time as well):

\begin{equation} \label{E:QHJ 2}
\frac{ {( \nabla S )}^2 }{2m}+V(x)+Q+\frac{\partial S}{\partial
t}=0
\end{equation}

\begin{equation} \label{E:Contin Eq 2}
\frac{\partial R^2}{\partial t}+ \nabla . ( R^2 \frac{\nabla S}{m}
)=0.
\end{equation}

Needless to say that the Eq. \eqref{E:Var principle2} is in fact a
special case of the equation \eqref{E:Int R2 H+St}. If $R$ is not
an explicit function of time, and $S$ is written in the form of
\eqref{E:S=W-Et}, the time integration in equation \eqref{E:Int R2
H+St} reduces to a multiplying constant, and therefore the
equation \eqref{E:Int R2 H+St} leads to the earlier result, i.e.,
the
equation \eqref{E:Var principle2} with $\lambda = E$.\\

\section{Relation with the usual Bohmian mechanics}
\nin In the previous section, We obtained the Bohmian equations of
quantum theory from simple considerations, without starting from
Schr\"odinger equation. Mathematically, the set of Eqs.
\eqref{E:QHJ 2} and \eqref{E:Contin Eq 2} along with the condition

\begin{equation}\label{E:W-S Condition}
    \oint \nabla S . dx = nh
\end{equation}

\noindent is equivalent to the Schr\"odinger equation for $\psi =
R e^{iS/ \hbar}$. The condition \eqref{E:W-S Condition} means that
the phase of wave function is unique and thus the wave function is
single-valued. Without this condition one can not consider $S$ as
phase of a wave function. Establishing this condition is
equivalent to appealing to the wave function, and denying it means
denying wave function in the theory.

In our approach, there is no need and no reason for imposing the
condition \eqref{E:W-S Condition}. This means that we do not
appeal to the wave function. This is a major difference between
quasi-Newtonian approach and usual Bohmian mechanics. Indeed, in
the quasi-Newtonian approach $S$ is the same as classical
Hamilton-Jacobi principal function: there is no uniqueness
condition \eqref{E:W-S Condition} on $S$.

Another difference between quasi-Newtonian approach and usual
Bohmian mechanics is connected with the denial of the condition
\eqref{E:W-S Condition}. When we do not need to consider the $S$
as phase of a wave function there is no need to consider $R$ as
amplitude of a wave function. Therefore, there is no need to
consider $R$ as a positive-definite function. We know from usual
Bohmian mechanics and also from our approach that $R$ appears in
the forms of $R^2$ or $\nabla^2 R / R$, thus negative values for
$R$ is not a problem. In a subsequent paper on 'quantization', we
shall show that the imposition of the uniqueness condition on $S$
and the positive-definiteness condition on $R$ are not necessary
for solving quantum problems.

\section{Conclusion}
\nin As we observed in this paper, one can start from a
quasi-Newtonian approach and get the mathematical form of quantum
potential by minimizing the total energy of ensemble, without
appealing to the Schr\"odinger equation and wave function. This
approach yields that the non-trivial quantum potential necessarily
is in the Bohmian form. After the derivation of the mathematical
form of quantum potential, if we impose the extra uniqueness
condition on $S$ (which is not necessary in quasi-Newtonian
approach), one can obtain the Schr\"odinger equation. This means
that one can consider the Bohmian quantum potential as the basis
of the Schr\"odinger equation rather than being a consequence of
it. In this picture, the quantum potential is the fundamental
concept of quantum theory, because it provides for the classical
mechanics
the possibility of existing non-classical effects.\\

\appendix
\section{} \label{A:Appendix}

\nin It is simple to prove that the equation \eqref{E:Var
principle2} with Bohmian quantum potential \eqref{E:BQP Def} is an
integral form of the time-independent Schr\"odinger equation. For
stationary states we have

\begin{equation*}
    \psi^\star \hat{H} \psi=R^2 \Bigg\{ \frac{( \nabla S )^2}{2m}+V
    -\frac{\hbar^2}{2m}\frac{\nabla^2 R}{R} \Bigg\}
\end{equation*}

Using the quantum potential $Q=AR^{-1} \nabla^2 R$ with
$A=-\hbar^2 /2m$, equation \eqref{E:Var principle2} becomes

\begin{equation*}
    \delta \int \big\{ R^2 H - \lambda R^2 \big\} \, d^3x=\delta
    \int \psi^\star \big\{ \hat{H} \psi - \lambda \psi \big\} \,
    d^3x=0
\end{equation*}

\noindent Variation with respect to $\psi^\star$ yields

\begin{equation*}
    \hat{H} \psi = \lambda \psi
\end{equation*}

\noindent that is time-independent Schr\"odinger equation with
eigenvalue
$\lambda$.\\

\end{document}